# A First Look at the JWST MIRI/LRS Phase Curve of WASP-43b

## Authors

Taylor J. Bell (BAERI), Laura Kreidberg (MPIA), Sarah Kendrew (ESA/STScI), Jacob Bean (U. Chicago), Nicolas Crouzet (Leiden University), Elsa Ducrot (CEA), Achrène Dyrek (CEA Paris), Peter Gao (Carnegie), Pierre-Olivier Lagage (CEA), and Julianne I. Moses (SSI), on behalf of the Transiting Exoplanet Community Early Release Science Team

## Executive Summary

We observed a full-orbit phase curve of the hot Jupiter WASP-43b with MIRI/LRS as part of the Transiting Exoplanet Community Early Release Science Program. Here we report preliminary findings from the team's MIRI Working Group. Overall we find that MIRI's performance for phase curve observations is excellent, with a few minor caveats.

**The key takeaways for Cycle 2 planning with MIRI/LRS are:** (1) long-duration (>24 hour) observations have now been successfully executed; (2) for phase curves, we recommend including a one-hour burn-in period prior to taking science data to mitigate the effects of the ramp systematic; and (3) we do not yet recommend partial phase curve observations.
In addition, we also find that:

- The position of the spectrum on the detector is stable to within 0.03 pixels over the full 26.5-hour observation.
- As with other LRS observations, the light curves typically show a systematic downward ramp that is strongest for the first 30 minutes, but continues to decay for hours.
- From 10.6–11.8 μm, the ramp effect has remarkably different behavior. Instead of a steep downward ramp, these wavelengths show a stronger, more gradual upward ramp. This is not the case for all LRS observations, but it is currently unpredictable if/when it will happen.
- The anomalous upward ramp behavior occurs in a "shadowed" region of the detector, with a different illumination history from other regions.
- After trimming the integrations most affected by the initial ramps and correcting the remaining systematics with analytic models, we obtain residuals to the light-curve fits that are typically within 25% of the photon noise limit for 0.5 μm spectroscopic bins. Non-linearity correction is not a significant source of additional noise for WASP-43 (though it may be an issue for brighter targets).
- The gain value of 5.5 electrons/DN currently on CRDS and JDox is known to be incorrect, and the current best estimate for the gain is ~3.1 electrons/DN. This is an empirically derived value for this dataset; the full implementation of the gain for noise calculations includes a wavelength-dependency and this work is in progress at STScI.

## Context

This document is a summary of the results obtained so far for the MIRI/LRS phase curve observation of WASP-43b, taken as part of the Transiting Exoplanet Community Early Release Science Program (DD-ERS 1366). The phase curve was observed relatively late in Cycle 1, but to help the community prepare for the Cycle 2 deadline, we are providing a snapshot here of the data reduction and analysis. For the most up-to-date information on data from this program, contact MIRI Working Group co-leads Laura Kreidberg and Nicolas Crouzet. We also recommend the submitted paper on MIRI's time series performance from Bouwman et al. (2022).

## Observation

We observed a full-orbit phase curve of the hot Jupiter WASP-43b with JWST MIRI/LRS on 1 December 2022. The observation consisted of continuous low-resolution time-series spectroscopy over 26.5 hours with a total of 9216 integrations each consisting of 64 groups.

## Instrument Stability

Instrument stability is key for phase curve observations. As an initial diagnostic of stability, we measured the center of the spectrum's spatial profile and the width of the profile over time. The position of the spectrum in the spatial direction is stable to within 0.0036 pixels RMS (0.027 pixels peak-to-peak) over the entire observation. The spatial PSF width is also constant to within 0.00069 pixels RMS (0.0084 pixels peak-to-peak), following a sharp increase of 0.022 pixels during the first 600 integrations. Movement in the spatial position of the spectrum appears to introduce noise at the ~1 ppm level (in 0.5 μm bins) while variations in the spatial profile introduce noise at the ~10 ppm level (in 0.5 μm bins). We do not see any obvious shifts in position or profile width corresponding to high gain antenna (HGA) moves; however, our team is still investigating this possibility. We are also planning to measure position shifts in the dispersion direction, which have been found to correlate with long-term drift in the measured flux for other data sets (GO 1803, private comm.).

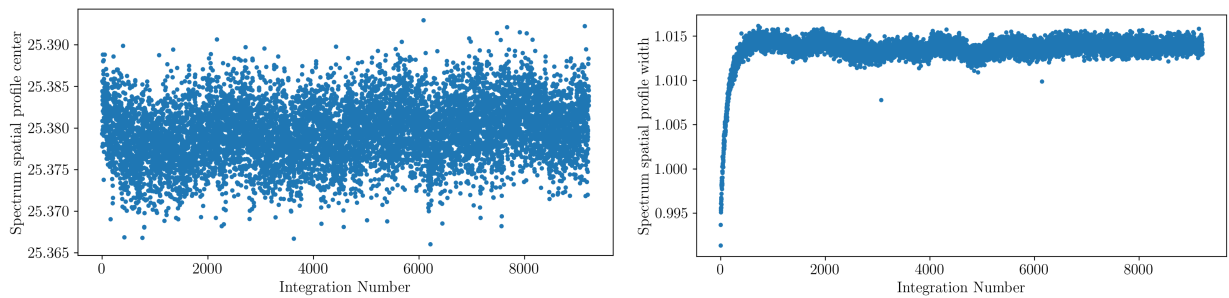

**Figure 1:** *Left*: Variations in the spatial position of the star throughout the observations. The profile center is computed by summing the spectra along the dispersion direction and then fitting a Gaussian to the resulting profile. *Right*: Variations in the spatial PSF width of the star throughout the observations. There is initially a strong ramp upwards, and then the PSF becomes quite stable. The PSF profile is computed using the same method as the center position and the width is the standard deviation from the Gaussian fit.

## Precision Compared to Expectations

The gain value of 5.5 electrons/DN currently on CRDS and JDox is known to be incorrect, and the current best estimate for the gain is ~3.1 electrons/DN (private comm.). The gain has been found to be wavelength dependent, and this is captured in the Photon Conversion Efficiency (PCE) that is included in the latest version of the JWST ETC engine Pandeia, which is also used by PandExo. The PCE reference file can be retrieved from the Pandeia data package that is available for download from this location. The value of 3.1 electrons/DN is consistent with the wavelength-variable gain in the LRS wavelength region (private comm.). New reference files for the JWST calibration pipeline reflecting these findings are under development at STScI; these will improve the accuracy of the pipeline noise calculations. For the rest of this document though, we will assume a constant gain of 3.1 electrons/DN.

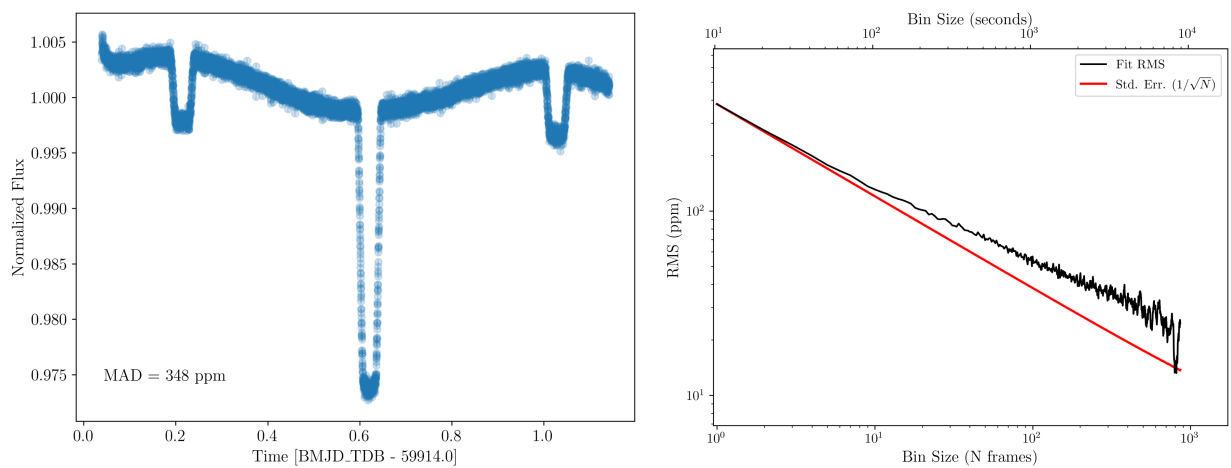

**Figure 2:** *Left*: The raw white lightcurve (5–12 μm) of WASP-43b with sigma clipping performed to remove rare outliers. Overall, besides an initial downward ramp at the start of the observations and a small downward linear trend over the whole observation, very little systematic noise appears present by eye. *Right*: The Allan variance plot after fitting the white lightcurve and removing the first 600 integrations which were especially affected by the ramp behavior.

After fitting astrophysical and systematic models to the data, we find that the residuals are typically ≲25% above the photon limit for 0.5 μm bins. Depending on the reduction, residuals range from 15–50% above the photon limit for white-light (5–12 μm) binning. The Median Absolute Deviation (MAD) of the white lightcurve is ~15% above the photon limit which gives an estimation of the high-frequency noise level that is model independent. The Allan variance plot (Allan 1966) for the white-light residuals in the right panel of Figure 2 shows there is a mild amount of red noise remaining after fitting the data, but it is likely that the majority of this is driven by a) the messy "shadowed region" discussed in the Instrument Systematics section below and b) the fact that the amplitude of the astrophysical signal changes across the full MIRI/LRS wavelength range. In individual 0.5 μm bins, the Allan variance plots show minimal evidence for residual red noise for wavelengths outside of the "shadowed region" (shorter than 10.6 μm).

We also note that the WASP-43b data do not show increased noise at short wavelengths, in contrast to the MIRI/LRS TSO commissioning observations of L168-9b (see Fig. 3). The difference is likely due to inadequate correction of non-linearity in the detector response for L168-9. L168-9 is brighter than WASP-43b, and has a relatively small number of groups per integration (9 versus 64 for WASP-43), which makes the non-linearity correction more challenging. For targets as faint as WASP-43 (K mag = 9.27), detector non-linearity is not a significant source of error, whereas it may be an issue for brighter targets like L168-9 (K mag = 7.08).

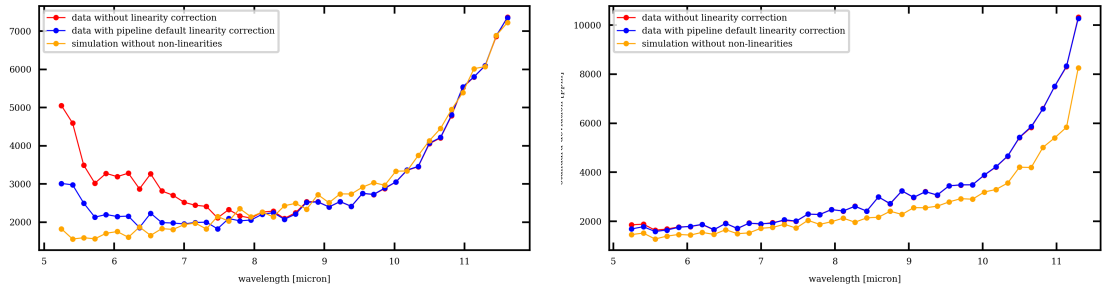

**Figure 3:** *Left*: RMS of the light curves after spectral extraction as a function of the wavelength for the observation of L168-9 from real data and simulated data (Dyrek et al., in preparation). The excess noise in the red curve comes from non-linearities at the end of each ramp. Correction of these non-linearities with the pipeline (blue curve) allows to decrease the noise but not entirely compared to what we should expect in a photon noise-limited scenario (orange curve). *Right*: RMS of the light curves as a function of the wavelength for WASP-43. No excess noise is observed at short wavelengths in this case and the linearity correction is not a significant source of noise (see red and orange curves).

## Instrument Systematics

As with other MIRI/LRS observations, our observations show a significant downward ramp at the start of the observations; this effect is discussed more thoroughly in [Bouwman et al. (2022)](#). In summary, there is a strong downward ramp that is especially strong for ~30 minutes or so but continues to decay for hours. As with [Bouwman et al. (2022)](#), the ramp can be modeled with a combination of exponentially decaying functions; however, for our longer duration observations we are sensitive to longer timescale ramp effects, so we require at least 3 exponentially decaying models to fit all integrations well with timescales in the ballpark of minutes, tens of minutes, and hours. If instead the first 600 integrations are removed (which showed the strongest PSF-width changes and the strongest flux changes), the shortest of the timescales becomes unimportant and the ramp behavior can be well-fit with 1–2 ramp timescales. In addition to the downward ramp behavior, there is also a downward linear slope to the observations. While the ramp behavior can be fairly well modeled with exponential functions, ***we recommend adding ~1 hour of padding to the start of MIRI/LRS phase curve observations*** which are especially affected by this kind of systematic curvature. Meanwhile, for transit and eclipse observations, the astrophysical shapes and timescales are far more decoupled from the ramp shape and timescale, so less than one hour of padding may suffice, as was seen for the transit observations of [Bouwman et al. (2022)](#).

Additionally, our observations show an, as yet unpublished, effect which degrades the quality of our spectrum from 10.6–11.8 μm (pixel rows 156–220). At these specific wavelengths, the observations show a remarkably different behavior than at shorter or longer wavelengths (see Figures [4](#) and [5](#)). Instead of a steep downward ramp effect, at these wavelengths there is a stronger and much more gradual upward ramp that slightly overshoots and then decays back down. This abrupt change was not seen in the MIRI/LRS TSO commissioning observations of L168-9b ([Bouwman et al. 2022](#)), and to our knowledge has been seen in only one other observation: the GTO transit observation of WASP-80b (GTO 1177; priv. comm.). Strangely, the same effect was not seen in the eclipse observation of WASP-80b (GTO 1177; priv. comm.) that was collected only 36 hours after the affected transit observation despite the fact that those eclipse observations used the exact same observing procedure as the affected transit observations.

As a result, it appears that this effect is not driven by the local background or nearby contaminants. As shown in the right panel of Figure 4, the affected pixels are in the "shadowed" region behind a portion of the focal plane mask between the Lyot and nearby 4 Quadrant Phase Mask regions on the detector. The "shadowed" region is not exposed to the sky when using imaging modes, including the WASP-43b target acquisition image shown in the foreground of the right panel of Figure 4 or the full-frame imaging mode observations shown in the background of that figure. Additionally, there is no shutter or pickoff mirror that obstructs the detector while the instrument is not in use; instead it is exposed to the sky at all times and is continually reset with a cadence of ~2.77 seconds (a full frame read time).

It is currently our suspicion that the sharp change in ramp behavior in 10.6–11.8 μm region compared to the non-shadowed parts of the subarray is caused by the prior illumination of the detector, while idling, or in the course of the target acquisition sequence. The flux falling on the illuminated portions of the subarray while idling is determined by the filter wheel position at the end of the previous MIRI imaging observation in the observing schedule. If this hypothesis is true, this would suggest the effect may easily be mitigated in the future through a modification to the telescope operation software. We emphasize that this effect and its causes are still under investigation by the MIRI team and in the community, and at this point there is no certain way of predicting whether an observation will show this effect. Some submitted analyses from the ERS team appear to have had moderate success in detrending the affected "shadowed" region, but it is not yet clear how robustly this can be done. However, for the other commissioning, GTO, and GO observations that we know do not show this effect, this wavelength region shows no unusual behavior and appears to be completely safe to use.

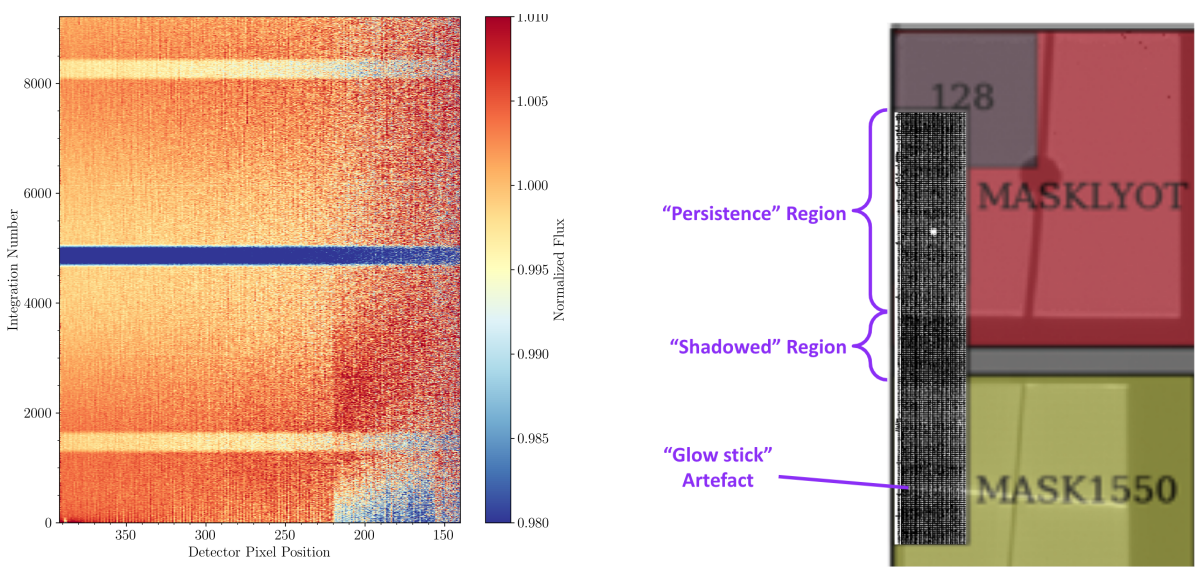

**Figure 4:** *Left*: A two-dimensional lightcurve showing integration number on the y-axis and pixel index along the x-axis (mirrored to have smaller indices which correspond to longer wavelengths on the right). The transit can clearly be seen as the dark blue band in the middle of the observations, and the two observed eclipses can clearly be seen as the yellow bands near the top and bottom of the figure. The initial downward ramp in the data can be seen in the red coloration along the bottom left of the figure, while the starkly different ramp behavior in the "shadowed" region can be seen in the blue and then red bands between pixel indices 156–220. The phase variations are visible in the redder coloration near both eclipses and yellower coloration near transit. ***Right***: A demonstration of the "shadowed" region that is believed to cause sharp change in ramp behavior at longer wavelengths. In the foreground is the target acquisition image for these observations; the star is clearly visible, as is the ["glow stick" artifact](#), and the region labeled as "shadowed" shows significantly less background flux compared to the rest of the detector. In the midground, an overlay of the location of the different subarrays is shown in red and yellow. In the background, a full-frame imaging observation is shown which demonstrates that the same "shadowed" region exists where there are pixels not exposed to the sky.

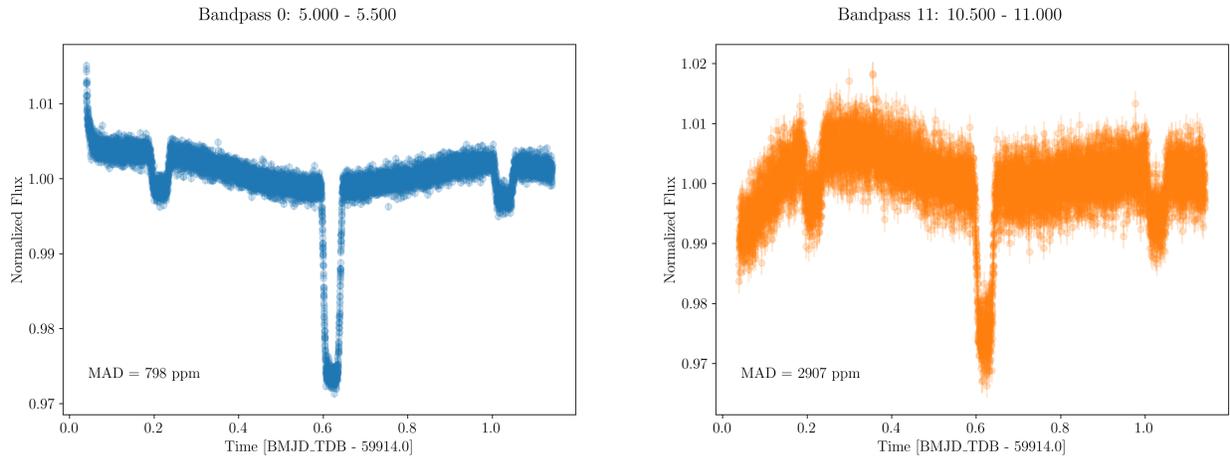

**Figure 5:** ***Left***: The uncorrected phase curve for WASP-43b binned across 5.0–5.5 μm with some sigma clipping performed to remove rare outliers. The strong downward ramp is seen at these wavelengths, but the amplitude of the ramp decreases with increasing wavelength. ***Right***: The same figure for 10.5–11.0 μm which lies within the "shadowed" region and shows a remarkably different ramp behavior.

## Conclusions

Overall, we find that MIRI/LRS can successfully measure thermal phase curves for exoplanets, with precision within 25% of the photon limit over a >24-hour time span. Non-linearity correction is not a significant source of error for the WASP-43b data (but it is an issue for brighter targets, e.g. L 168-9; [Bouwman et al. 2022](#)). The light curves do have some instrumental systematics, most notably a ramp effect that is strongest during the first 30 minutes of the observation, but continues for several hours. The morphology of the ramp changes sharply over the wavelength range 10.6–11.8 μm; this corresponds to a "shadowed" region of the detector. We recommend investigating whether the anomalous ramp behavior is caused by a particular filter left in the optical path from the preceding observation.

To mitigate the systematic ramp effect for phase curve observations, we recommend including one hour of burn-in time prior to the start of science observations. For shorter transit and eclipse observations, it is easier to decouple the ramp from the astrophysical signal, so for these cases less baseline is likely needed (see [Bouwman et al. 2022](#)).

At this early stage we do not recommend partial phase curve observations with MIRI/LRS. Further assessments of the ramp systematic and the visit-to-visit stability are needed to determine whether partial phase curves are feasible.